\documentclass[twoside,twocolumn,9pt]{article}
\usepackage{pdfpages}
\usepackage{extsizes}
\usepackage{changepage}
\usepackage{multirow}
\usepackage[super,sort&compress,comma]{natbib} 
\usepackage[version=3]{mhchem}
\usepackage[left=1.5cm, right=1.5cm, top=1.785cm, bottom=2.0cm]{geometry}
\usepackage{balance}
\usepackage{mathptmx}
\usepackage{sectsty}
\usepackage{graphicx} 
\usepackage{lastpage}
\usepackage{multirow}
\usepackage[format=plain,justification=justified,singlelinecheck=false,font={stretch=1.125,small,sf},labelfont=bf,labelsep=space]{caption}
\usepackage{float}
\usepackage{fancyhdr}
\usepackage{fnpos}
\usepackage[english]{babel}
\addto{\captionsenglish}{%
  \renewcommand{\refname}{Notes and references}
}
\usepackage{array}
\usepackage{droidsans}
\usepackage{charter}
\usepackage[T1]{fontenc}
\usepackage{setspace}
\usepackage[compact]{titlesec}
\usepackage{hyperref}
\usepackage{physics}

\newcommand{\RE}{\mathrm{Re}}
\newcommand{\IM}{\mathrm{Im}}

\usepackage{epstopdf}%This line makes .eps figures into .pdf - please comment out if not required.

\definecolor{cream}{RGB}{222,217,201}

\begin{document}

\pagestyle{fancy}
\thispagestyle{plain}
\fancypagestyle{plain}{
%%%HEADER%%%
\renewcommand{\headrulewidth}{0pt}
}
%%%END OF HEADER%%%

%%%PAGE SETUP - Please do not change any commands within this section%%%
\makeFNbottom
\makeatletter
\renewcommand\LARGE{\@setfontsize\LARGE{15pt}{17}}
\renewcommand\Large{\@setfontsize\Large{12pt}{14}}
\renewcommand\large{\@setfontsize\large{10pt}{12}}
\renewcommand\footnotesize{\@setfontsize\footnotesize{7pt}{10}}
\makeatother

\renewcommand{\thefootnote}{\fnsymbol{footnote}}
\renewcommand\footnoterule{\vspace*{1pt}% 
\color{cream}\hrule width 3.5in height 0.4pt \color{black}\vspace*{5pt}} 
\setcounter{secnumdepth}{5}

\makeatletter 
\renewcommand\@biblabel[1]{#1}            
\renewcommand\@makefntext[1]% 
{\noindent\makebox[0pt][r]{\@thefnmark\,}#1}
\makeatother 
\renewcommand{\figurename}{\small{Fig.}~}
\sectionfont{\sffamily\Large}
\subsectionfont{\normalsize}
\subsubsectionfont{\bf}
\setstretch{1.125} %In particular, please do not alter this line.
\setlength{\skip\footins}{0.8cm}
\setlength{\footnotesep}{0.25cm}
\setlength{\jot}{10pt}
\titlespacing*{\section}{0pt}{4pt}{4pt}
\titlespacing*{\subsection}{0pt}{15pt}{1pt}
%%%END OF PAGE SETUP%%%

%%%FOOTER%%%
\fancyfoot{}
\fancyfoot[LO,RE]{\vspace{-7.1pt}\includegraphics[height=9pt]{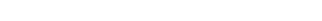}}
\fancyfoot[CO]{\vspace{-7.1pt}\hspace{13.2cm}\includegraphics{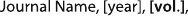}}
\fancyfoot[CE]{\vspace{-7.2pt}\hspace{-14.2cm}\includegraphics{head_foot/RF}}
\fancyfoot[RO]{\footnotesize{\sffamily{1--\pageref{LastPage} ~\textbar  \hspace{2pt}\thepage}}}
\fancyfoot[LE]{\footnotesize{\sffamily{\thepage~\textbar\hspace{3.45cm} 1--\pageref{LastPage}}}}
\fancyhead{}
\renewcommand{\headrulewidth}{0pt} 
\renewcommand{\footrulewidth}{0pt}
\setlength{\arrayrulewidth}{1pt}
\setlength{\columnsep}{6.5mm}
\setlength\bibsep{1pt}
%%%END OF FOOTER%%%

%%%FIGURE SETUP - please do not change any commands within this section%%%
\makeatletter 
\newlength{\figrulesep} 
\setlength{\figrulesep}{0.5\textfloatsep} 

\newcommand{\topfigrule}{\vspace*{-1pt}% 
\noindent{\color{cream}\rule[-\figrulesep]{\columnwidth}{1.5pt}} }

\newcommand{\botfigrule}{\vspace*{-2pt}% 
\noindent{\color{cream}\rule[\figrulesep]{\columnwidth}{1.5pt}} }

\newcommand{\dblfigrule}{\vspace*{-1pt}% 
\noindent{\color{cream}\rule[-\figrulesep]{\textwidth}{1.5pt}} }

\newcommand{\BW}[1]{\noindent\textcolor{red}{ #1}}
\newcommand{\BWcomm}[1]{\noindent\textcolor{blue}{ #1}}

\makeatother
%%%END OF FIGURE SETUP%%%

%%%TITLE, AUTHORS AND ABSTRACT%%%
\twocolumn[
  \begin{@twocolumnfalse}
{\includegraphics[height=30pt]{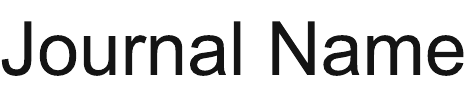}\hfill\raisebox{0pt}[0pt][0pt]{\includegraphics[height=55pt]{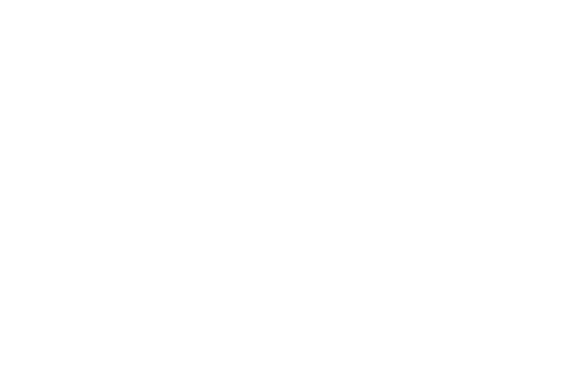}}\\[1ex]
\includegraphics[width=18.5cm]{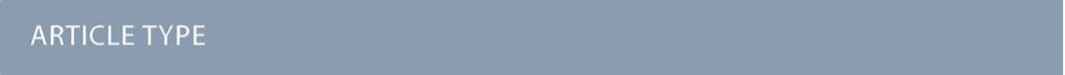}}\par
\vspace{1em}
\sffamily
\begin{tabular}{m{4.5cm} p{13.5cm} }

\includegraphics{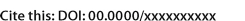} & \noindent\LARGE{\textbf{Beyond real: Alternative unitary cluster Jastrow models for 
%variational quantum calculations of molecules$^\dag$
molecular electronic structure calculations on near-term quantum computers$^\dag$}} \\%Article title goes here instead of the text "This is the title"
\vspace{0.3cm} & \vspace{0.3cm} \\

 & \noindent\large{Nikolay V. Tkachenko,\textit{$^{a,b,c}$}$^{\ast}$ 
Hang Ren,\textit{$^{a}$} 
Wendy M. Billings,\textit{$^{a}$} 
 %Andrea Rodriguez-Blanco,\textit{$^{a}$} 
Rebecca Tomann,\textit{$^{a}$} 
K. Birgitta Whaley\textit{$^{a,d}$}$^{\ast}$} and Martin Head-Gordon,\textit{$^{a,c,d}$}$^{\ast}$ 
\\%Author names go here instead of "Full name", etc.

\includegraphics{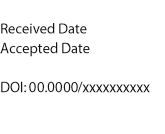} & \noindent\normalsize{Near-term quantum devices require 
%shallow yet expressive 
wavefunction ansätze that are expressive while also of shallow circuit depth in order to both accurately and efficiently simulate 
molecular electronic structure. While unitary coupled cluster (e.g., UCCSD) has become a standard, 
the high gate count associated with the implementation of this limits its feasibility on noisy intermediate-scale quantum (NISQ) hardware. K-fold unitary cluster Jastrow (uCJ) 
ansätze mitigate this challenge by providing $O(kN^2)$ circuit scaling and favorable linear depth circuit implementation. Previous work has
focused on the real orbital-rotation (Re-uCJ) variant of uCJ, which allows an exact (Trotter-free) implementation. Here 
we extend and generalize the $k$-fold uCJ framework by introducing two new variants, Im-uCJ and g-uCJ, which incorporate imaginary 
and fully complex orbital rotation operators, respectively. Similar to Re-uCJ, both of the new variants achieve quadratic gate-count scaling. Our results focus on the simplest $k=1$ model, and show that the uCJ models frequently maintain 
energy errors within chemical accuracy. Both g-uCJ and Im-uCJ are more expressive in terms of capturing electron correlation and are also more accurate than the earlier Re-uCJ ansatz. We further show that Im-uCJ and g-uCJ circuits can also be implemented exactly, without any Trotter decomposition. Numerical tests using $k=1$ on \ce{H2}, \ce{H3+}, \ce{Be_2}, \ce{C2H4}, \ce{C2H6} and \ce{C6H6} in various basis sets confirm the practical feasibility of these shallow Jastrow-based ansätze for applications on near-term quantum hardware.}

\end{tabular}

 \end{@twocolumnfalse} \vspace{0.6cm}

  ]
%%%END OF TITLE, AUTHORS AND ABSTRACT%%%

%%%FONT SETUP - please do not change any commands within this section
\renewcommand*\rmdefault{bch}\normalfont\upshape
\rmfamily
\section*{}
\vspace{-1cm}

%%%FOOTNOTES%%%

\footnotetext{\textit{$^{a}$~Department of Chemistry, University of California, Berkeley, CA 94720, USA.}}
\footnotetext{\textit{$^{b}$~Materials Sciences Division, Lawrence Berkeley National Laboratory, Berkeley, CA 94720, USA.}}
\footnotetext{\textit{$^{c}$~Institute for Decarbonization Materials, University of California, Berkeley, CA 94720, USA.}}
\footnotetext{\textit{$^{d}$~Chemical Sciences Division, Lawrence Berkeley National Laboratory, Berkeley, CA 94720, USA.}}
\footnotetext{$^{\ast}$~E-mail: nikolaytkachenko@berkeley.edu; whaley@berkeley.edu; m\_headgordon@berkeley.edu.}
%Please use \dag to cite the ESI in the main text of the article.
%If you article does not have ESI please remove the the \dag symbol from the title and the footnotetext below.
\footnotetext{\dag~Supplementary Information available: [details of any supplementary information available should be included here]. See DOI: 00.0000/00000000.}
%additional addresses can be cited as above using the lower-case letters, c, d, e... If all authors are from the same address, no letter is required

%%%END OF FOOTNOTES%%%

%%%MAIN TEXT%%%%
\section{Introduction}

Electronic structure simulations represent one of the most promising scientific application areas where quantum computers have the potential to outperform classical methods. Since ab initio calculations play an essential role in various fields from catalysis to drug discovery, the significant advantages that quantum computers are expected to provide relative to classical approaches would open new possibilities for the prediction and analysis of complex systems.\cite{Cao2019, McArdle2020} In molecules lacking strong static electron correlations, methods such as Density Functional Theory (DFT)\cite{Kohn1965, Mardirossian2017} or wave function-based techniques such as coupled cluster (CC) methods \cite{Cizek1966, Shavitt2009, Christiansen1995, Koch1997, Raghavachari1989} often produce usefully accurate results, and have been successfully employed for decades. Nevertheless, both approaches have inherent drawbacks. On the one hand, the accuracy of DFT typically depends on the functional chosen, as well as the system under investigation, and can further be constrained by self-interaction errors. On the other hand, the most precise wave function-based methods are computationally expensive, limiting their use to small or moderately sized systems. Furthermore, in the case of strongly correlated systems, despite their accuracy, even the most advanced classical algorithms based on complete active space (CAS) methods,\cite{Angeli2001a,Angeli2001b,Angeli2002a,Angeli2002b} or the density matrix renormalization group (DMRG),\cite{Chan2002,Chan2004,Ghosh2008} or selected configuration interaction\cite{Levine2020} face unfavorable exponential scaling with molecule size (measured, e.g., by the number of atoms, electrons, spatial or spin orbitals), thereby restricting their application to small systems.

A wide variety of algorithms have been proposed for efficient electronic structure calculations on quantum computers.\cite{McArdle2020,Cao2019,Dalzell2023,Huang2023, Su2021}
While the quantum phase estimation (QPE) algorithm\cite{abrams1999quantum} offers
a scalable pathway for addressing the electronic structure problem,\cite{aspuru2005simulated} the high circuit costs of this algorithm and variants on it\cite{QMEGS} appear to require fault-tolerant quantum hardware.\cite{vonBurg2021,Bauman2021,Wecker2014}
The majority of algorithms proposed for the simulation of electronic structure on near-term noisy, intermediate scale quantum (NISQ) devices fall either within the category of variational optimization algorithms\cite{Peruzzo2014, McClean2016, Kandala2017, Higgott2019, Sim2018} or lie within the broad class of quantum subspace diagonalization (QSD)\cite{Huggins2020, Parrish2019, Seki2021, Stair2020} algorithms. In the variational optimization category, the variational quantum eigensolver (VQE)\cite{Peruzzo2014} is perhaps the most well-known. VQE uses a quantum computer to prepare trial wavefunctions and a classical computer to optimize the parameters of these wavefunctions. While initial studies indicated significant promise for ground-state energy calculations,\cite{Peruzzo2014, McClean2016} subsequent work showed that the classical optimization component is non-trivial, due to the frequent occurrence of barren plateaux that arise from shallow or even flat energy landscapes.\cite{McClean2018, Wang2021, Cerezo2021a, Arrasmith2021} Several extensions and modifications of VQE have since been developed to improve its accuracy and applicability.\cite{Lee2018, Wecker2015, Barkoutsos2018, Wiersema2020, Gard2020, Romero2018, Mizukami2020, Tkachenko2021, Grimsley2019, Tang2021} For example, PermVQE\cite{Tkachenko2021} introduces correlation-informed qubit permutation to the optimization process, allowing the accuracy of energy predictions to be improved without growing the circuit depth. Another example is ADAPT-VQE\cite{Grimsley2019}, which optimizes construction of the quantum circuit by selectively adding operators that provide the greatest reduction in energy at each iteration. The more recent qubit-ADAPT-VQE\cite{Tang2021} further enhances the performance of this adaptive approach by optimizing the operator selection at the qubit-operator level, thereby reducing the number of quantum gates required for efficient simulations.

An alternative class of quantum algorithms for calculating electronic energies on NISQ or near-term hardware is provided by subspace diagonalization algorithms. These focus on capturing electron correlation by expanding the state basis. These are hybrid algorithms that generally involve classical postprocessing of a Hamiltonian eigenvalue problem with the matrix elements evaluated on a quantum processor. These algorithms can be further classified according to the methods used to define the basis states, which may or may not be orthogonal. For instance, the Quantum Subspace Expansion (QSE) builds on VQE outputs by constructing an expanded subspace from the original VQE solution, e.g., $\hat{a}_p^\dag\hat{a}_q \ket{\Psi_{VQE}}$.~\cite{McClean2017, Colless2018} It is also possible to achieve provable convergence with respect to growth of the subspace when the set of expanded states forms a Krylov basis that is generated by the repeated application of the matrix of interest to an initial guess vector~\cite{Cortes2022}. Several algorithms have been proposed along this direction, including quantum filter diagonalization~\cite{Parrish2019}, quantum Lanczos~\cite{Motta2020}, and quantum Davidson methods~\cite{Tkachenko2024}. 
%Another variant 
A different form of subspace diagonalization is provided by the non-orthogonal quantum eigensolver (NOQE),\cite{Baek2023} which is a multi-reference method for systems with both strong and weak electronic correlations that offers both algorithmic and practical quantum advantages, and has recently been shown to provide a complexity theoretic quantum speedup for such systems.\cite{leimkuhler2025}

The work we report here focuses on the adaptations needed to successfully use cluster wavefunctions for systems with both strong and weak electronic correlations. As such, it is relevant to both variational methods and the NOQE. Classical CC theory converges only slowly with rank for strongly correlated systems\cite{Lehtola2016_JCP_134110} due to both its nonvariational character and the nature of the cluster expansion\cite{Small2012_JCP_114103}. The variational issue is well solved via the VQE and unitary CC (UCC) theory.\cite{Evangelista2019_JCP_244112,anand2022quantum}. However, the large number of variational parameters, even at the lowest singles and doubles (i.e., quartic scaling amplitudes at the UCCSD level) which results in large circuit depths has motivated development of more compact alternatives.\cite{Lee2018,dallaire2019low,motta2021low} We 
focus here on the unitary cluster Jastrow approximation (uCJ),\cite{matsuzawa2020jastrow}, which builds on the well-known real-space Jastrow factors from quantum Monte Carlo, extended into Hilbert space to become cluster operators.\cite{neuscamman2013communication} The uCJ approach was used in the NOQE\cite{Baek2023} where the reduction in gate count was noted, and has also been successfully mapped onto aspects of both physics of strong correlation and qubit connectivity with a local uCJ extension.\cite{motta2023bridging,motta2024quantum}

In this work, we introduce and systematically explore several variants of the uCJ ansätze, analyzing their circuit depth, expressibility, and accuracy in VQE calculations. Specifically, we examine two new forms of uCJ correlator, one with imaginary orbital rotation operators (Im-uCJ) and one with complex orbital-rotation operators (g-uCJ). We show that similar to the previously introduced Re-uCJ ansatz,\cite{Baek2023} these new ansätze also reduce the gate count relative to the widely used UCCSD ansatz, making the uCJ correlators more suitable for near-term hardware. 
We demonstrate that both Im-uCJ and g-uCJ can achieve high accuracy, often surpassing both UCCSD and their real-rotation counterpart (Re-uCJ), while preserving a shallow circuit depth. 
We benchmark these uCJ ansätze on a series of small molecules, demonstrating that cluster Jastrow correlators can serve as a key step toward practical and
resource-efficient quantum algorithms for molecular electronic structure on NISQ and near-term devices.

\section{Theory}
\subsection{Formulations of unitary Cluster Jastrow ansätze}

While UCC ansätze, such as UCCSD, are frequently used in variational quantum algorithms, 
their circuit depths often become impractical on near-term quantum hardware. One path toward 
lowering this overhead is to move away from two-body operators and build wavefunction ansätze from simpler one-body terms. In this work, 
we therefore focus on Jastrow-style correlators, which use exponentials of one-electron operators and particle number operators.~\cite{matsuzawa2020jastrow} 
The k-fold uCJ ansatz\cite{matsuzawa2020jastrow} is expressed as
\begin{equation}
    \label{eq:uCJ}
    \ket{\Psi}=\prod_{i=1}^{k}e^{-\hat{K}_i}e^{\hat{J}_i}e^{\hat{K}_i}\ket{HF}
\end{equation}
where the operators $\hat{K}$ and $\hat{J}$ are defined as
\begin{equation}
    \label{eq:K}  \hat{K}=\sum_{pq,\sigma}K_{pq,\sigma\sigma}\hat{a}_{p,\sigma}^\dag\hat{a}_{q,\sigma}
\end{equation}
\begin{equation}
    \label{eq:J}
    \hat{J}=\sum_{pq,\sigma\tau}J_{pq,\sigma\tau}\hat{n}_{p,\sigma} \hat{n}_{q,\tau}
\end{equation}
In the present formulation of the $\hat{K}$ operator, we restrict orbital rotations to 
occur only within the same spin subspace. This choice constrains the number of variational 
parameters and simplifies the ansatz. However, lifting this restriction (allowing rotations connecting different spin states) 
increases the number of variational parameters (while still scaling quadratically with the number of spin orbitals) and can enhance expressibility. 
In this work, we primarily investigate the spin-preserving form, and will state explicitly when the spin-generalized form of $\hat{K}$ is used.

In equations \ref{eq:uCJ}-\ref{eq:J}, $p$ and $q$ refer to spatial molecular orbitals, $\sigma$ and $\tau$ represent spin polarization (either $\alpha$ or $\beta$), and $\ket{HF}$ is the mean-field restricted or unrestricted Hartree-Fock reference state. The parameter $k$ controls how many replicas of the Jastrow type correlators are included in the ansätz. If $k$ is not truncated, equation \ref{eq:uCJ} can be exact.\cite{Evangelista2019_JCP_244112,matsuzawa2020jastrow} Nevertheless, we shall limit ourselves to $k=1$ since this choice is simplest and cleanest, and we wish also to explore the extent to which this can provide a good balance between capturing a significant portion of electron correlations while maintaining a shallow enough ansatz to be interesting for near-term devices.
For notational simplicity, we will omit the $\sigma$ subscripts in $K_{pq,\sigma\sigma}$, while noting that the orbital rotations only couple orbitals of the same spin. To maintain unitarity in the uCJ ansätz, the coefficients $K_{pq}$ and $J_{pq,\sigma\tau}$ must satisfy specific conditions: the matrix \textbf{K} is required to be anti-Hermitian, while \textbf{J} must be purely imaginary and symmetric. 

Previous works\cite{matsuzawa2020jastrow,Baek2023,motta2023bridging} have explored the uCJ ansatz with real orbital rotations, which imposes the restriction that \textbf{K} is real and anti-Hermitian. We refer to this form as real uCJ (Re-uCJ). 
In this case, the effective form of the $\hat{K}$ operator can be written as:
\begin{equation}
    \label{eq:Re-uCJ}  \hat{K}_\mathrm{Re-uCJ}=\sum_{pq}K_{pq}(\hat{a}_{p}^\dag\hat{a}_{q} - \hat{a}_{q}^\dag\hat{a}_{p})
\end{equation}

However, the flexibility of the uCJ ansatz also allows additional choices that to our knowledge have not yet been explored. It is particularly interesting to explore whether different choices can yield significant improvements for truncation at $k=1$ that we impose throughout this work. 

The first alternative we consider is the use of the $\hat{K}$ operator with the opposite restriction to that imposed in Re-uCJ, i.e., restricting \textbf{K} to be imaginary.  We thereby obtain another form of the uCJ ansatz, which we shall refer to as imaginary uCJ (Im-uCJ):
\begin{equation}
    \label{eq:Im-uCJ}  \hat{K}_\mathrm{im-uCJ}=\sum_{pq}K_{pq}(\hat{a}_{p}^\dag\hat{a}_{q} + \hat{a}_{q}^\dag\hat{a}_{p})
\end{equation}
The second alternative is to consider the least restricted scenario for \textbf{K}, where it is allowed to be complex with both real and imaginary parts. This formulation provides the greatest flexibility for a given truncation of $k$, and therefore we refer to it as generalized uCJ (g-uCJ). The 
%form of the 
$\hat{K}$ operator for g-uCJ is written as:
\begin{equation}
    \label{eq:g-UCJ}  \small{\hat{K}_\mathrm{g-uCJ}=\sum_{pq}\left[\RE(K_{pq})(\hat{a}_{p}^\dag\hat{a}_{q} - \hat{a}_{q}^\dag\hat{a}_{p}) + i\cdot \IM(K_{pq})(\hat{a}_{p}^\dag\hat{a}_{q} + \hat{a}_{q}^\dag\hat{a}_{p})\right]}
\end{equation}

The overall scalability of the Jastrow ansatz is significantly more favorable than other UCC variants, since the number of terms in $e^{\hat{K}_i}$ and $e^{\hat{J}_i}$ scale as $O(N^2)$, where $N$ is the number of spin orbitals. This results in an overall $O(N^2)$ scaling of the uCJ ansatz, in contrast to the formal $O(N^4)$ scaling for a single Trotter step of a UCC ansatz with double excitations~\cite{UCC_for_QC_aspuruguzik}. Additionally, by applying a fermionic-to-spin transformation to $\hat{K}$ (e.g., the Jordan-Wigner (JW) transformation), we find that the restricted Re-uCJ and im-uCJ ansätze are both represented with a number of Pauli words that is smaller by a factor of two than that for the g-uCJ ansatz. Once the ansatz is defined, the parameters in the matrices \textbf{K} and \textbf{J} can be variationally optimized, either classically or in principle on a quantum computer. In this work, we test the performance of all three uCJ ansatz variants, Re-uCJ, Im-uCJ, and g-uCJ, with classical optimization of the parameters, as would be carried out, e.g., within a VQE approach.

\subsection{Implementation of the Im-uCJ and g-uCJ ansätze through Givens rotations}

Upon transforming the operators $\hat{K}$ and $\hat{J}$ from the fermionic to a qubit representation, the question arises of how to accurately represent the exponentials of these operators. In general, once $\hat{K}$ and $\hat{J}$ are mapped into qubit space, they may consist of Pauli words that do not commute with each other. A straightforward solution to this non-commutativity is to employ the approximate Trotter decomposition. However, for the specific case of the JW-transformation, the need for Trotter decomposition can be avoided. In the JW mapping, the exponentiation of the $\hat{J}$ operator is straightforward to handle, since the number operators are mapped to commuting $\hat{Z}$ Pauli operators, according to  
\begin{equation}
    \label{eq_12}  \hat{n}_p\hat{n}_q=\hat{a}_{p}^\dag\hat{a}_{p}\hat{a}_{q}^\dag\hat{a}_{q}=\frac{1}{4}\left(1-\hat{Z}_p\right)\left(1-\hat{Z}_q\right).
\end{equation}
The JW mapping then allows us to express the sum of these operators as an exact product of exponentials of individual terms.

On the other hand, the JW-transformed $\hat{K}$ operator generally includes non-commuting terms. Implementation of this operator presents a greater challenge, but can be realized by taking advantage of the demonstration that any real unitary orbital rotation operator can be implemented efficiently using Givens rotation operators.\cite{Kivlichan2018} Here we extend this approach to show that the exponentials $\exp({\hat{K}_\mathrm{Im-uCJ}})$  and $\exp({\hat{K}_\mathrm{g-uCJ}})$ can also be represented as consecutive applications of a generalized form of Givens rotation.

% Ref.~\citenum{Kivlichan2018} showed that any real orbital rotation unitary $\hat U(\mathbf{u})$ of the form: 
%\bwcomm{BW: Nikolay, I rewrote from here to eq. 18, please check the new version. Also see my question about the sign of the phases in the expression in the text line below eq. 13}
The analysis of Ref.~\citenum{Kivlichan2018} showed that any particle number-preserving rotation operator of the single-particle basis 
\begin{equation}
\hat{U}(\mathbf{u}) = \exp\left( \sum_{p,q} [\log \mathbf{u}]_{pq} \left( \hat{a}_p^\dagger \hat{a}_q - \hat{a}_q^\dagger \hat{a}_p \right) \right)
\label{eq:U_from_logu}
\end{equation}
where $\textbf u$ is a unitary matrix, can be efficiently decomposed into a sequence of $\binom{N}{2}$ real fermionic rotations of the form

\begin{equation}
    \label{eq_13}  
    \hat{R}_{pq}(\theta_{k})=\exp\left(\theta_{k}(\hat{a}_{p}^\dag\hat{a}_{q} - \hat{a}_{q}^\dag\hat{a}_{p})\right),
\end{equation}
The proof is based on the equivalence of application of the orbital rotation operator $\hat{R}_{pq}(\theta_k)$ to the unitary $\hat{U}(\mathbf{u})$ and rotations of matrix $\mathbf{u}$
% \bw{definition}
% \begin{equation}
% \mathbf{u} = \exp(\mathbf{K})
% \label{eq:U_small}
% \end{equation}
% the unitary $\hat{U}(\mathbf{u})$ in the equation \ref{eq:U_from_logu} transforms to the $e^{\hat{K}}$ operator. The decomposition procedure is based on the observation that the application of the orbital rotation operator $\hat{R}_{pq}(\theta_k)$ to the rotation unitary $\hat{U}(\mathbf{u})$ is equivalent to rotations of matrix $\mathbf{u}$, i.e.  
\begin{equation}
    \label{eq:extra}  
    \hat{R}_{pq}(\theta_{k})\hat{U}(\mathbf{u})=\hat{U}(\mathbf{r}_{pq}(\theta_k)\mathbf{u}),
\end{equation}
where $\mathbf{r}_{pq}(\theta_k)$ denotes a Givens rotation, represented by an N by N matrix (with N the number of spin-orbitals considered) of the form
\begin{equation}
\mathbf{r}_{pq}(\theta) =
\begin{bmatrix}
1 &  \dots      &  0      &  \dots    & 0   &  \dots   & 0       \\
\vdots  & \ddots &   \vdots     &        &   \vdots     &        & \vdots       \\
0  &  \dots       & \cos\theta &  \dots       & -\sin\theta &  \dots       &  0      \\
\vdots  &        &   \vdots      & \ddots &  \vdots       &        &   \vdots     \\
0  &   \dots      & \sin\theta &   \dots      & \cos\theta  &  \dots       & 0       \\
\vdots  &        &   \vdots     &        &    \vdots    & \ddots &  \vdots      \\
0  &   \dots      &    0    &   \dots      &    0    &   \dots      & 1      
\end{bmatrix},
\label{eq:givens_rotation}
\end{equation}
where the cosine terms occupy the $(p,p)$ and $(q,q)$ positions and oppositely signed sine terms occupy the $(p,q)$ and $(q,p)$ positions. 
%With the equation 
%We can now propose a decomposition scheme that is similar to QR-decomposition strategy. 
Identifying a sequence of Givens rotations $\mathbf{r}_{pq}(\theta_k)$ that diagonalize the matrix $\mathbf{u}$, which can be done by a QR-like decomposition as described in Ref.~\citenum{Kivlichan2018}, one has
\begin{equation}
\left( \prod_{k=1}^{M} \mathbf{r}_{pq}(\theta_k) \right) \mathbf{u} = \sum_{p=1}^{N} e^{i\phi_p} \ket{p} \bra{p}
\label{eq:qr_diag}
\end{equation}
and substituting this entire sequence in eq. \ref{eq:extra} leads to
\begin{equation}
\begin{aligned}
\left( \prod_{k=1}^{M} \hat{R}_{pq}(\theta_k) \right) \hat{U}(\mathbf{u}) &= \hat{U}\left(\left(\prod_{k=1}^{M} \mathbf{r}_{pq}(\theta_k) \right) \mathbf{u}\right) \\ & =  \hat{U}\left( \sum_{p=1}^{N} e^{i\phi_p} \ket{p} \bra{p}  \right) \\
& = \prod_{p=1}^{N} e^{i\phi_p \hat{n}_p},
\label{eq:qr_unitary_extra}
\end{aligned}
\end{equation}
from which the implementation of $\hat{U}(\mathbf{u})$ follows by
% Thus, to implement $\hat{U}(\mathbf{u})$, one applies 
applying the one-qubit phase gates $\prod_p e^{i\phi_p \hat{n}_p}$, followed by the inverse of the sequence of two-qubit rotations $\hat{R}_{pq}(-\theta_k)$:
\begin{equation}
\hat{U}(\textbf{u}) = \left( \prod_{k=M}^{1} \hat{R}_{pq}(-\theta_k) \right) \prod_{p=1}^{N} e^{i\phi_p \hat{n}_p}.
\label{eq:qr_unitary}
\end{equation}
This implementation can be directly applied to the Re-uCJ ansatz by choosing the unitary operator $\mathbf{u}$ as
%to represent the exponent of the $\hat{K}$ operator,
\begin{equation}
\mathbf{u} = \exp(\mathbf{K}),
\label{eq:U_small}
\end{equation}
so that the unitary $\hat{U}(\mathbf{u})$ in eqs. \ref{eq:U_from_logu} and \ref{eq:qr_unitary} correspond to the real $e^{\hat{K}}$ operator, which thus represents a real orbital rotation unitary and can be then implemented by eq. \ref{eq:qr_unitary}. 
However, the decomposition of Ref.~\citenum{Kivlichan2018} does not directly apply to the Im-uCJ and g-uCJ ansätze, which involve imaginary and complex orbital rotations, respectively, and therefore the corresponding matrices $\mathbf{u}$ are no longer real. 
To extend the method to be applicable with complex $\mathbf{u}$, we use a generalized Givens rotation of the form
\begin{equation}
\mathbf{r'}_{pq}(\theta,\phi) =
\begin{bmatrix}
1 &  \dots      &  0      &  \dots    & 0   &  \dots   & 0       \\
\vdots  & \ddots &   \vdots     &        &   \vdots     &        & \vdots       \\
0  &  \dots       & \cos\theta &  \dots       & -e^{-i\phi}\sin\theta &  \dots       &  0      \\
\vdots  &        &   \vdots      & \ddots &  \vdots       &        &   \vdots     \\
0  &   \dots      & e^{i\phi}\sin\theta &   \dots      & \cos\theta  &  \dots       & 0       \\
\vdots  &        &   \vdots     &        &    \vdots    & \ddots &  \vdots      \\
0  &   \dots      &    0    &   \dots      &    0    &   \dots      & 1      
\end{bmatrix},
\label{eq:givens_rotation2}
\end{equation}
together with the corresponding generalized fermionic rotations
\begin{equation}
    \label{eq:g-UCJ}  
    \hat{R}'_{pq}(\theta_{k},\phi_{k})=\exp\left[A(\hat{a}_{p}^\dag\hat{a}_{q} - \hat{a}_{q}^\dag\hat{a}_{p}) + iB(\hat{a}_{p}^\dag\hat{a}_{q} + \hat{a}_{q}^\dag\hat{a}_{p})\right],
\end{equation}
where the angles $\theta$ and $\phi$ are defined as $\theta=\sqrt{A^2+B^2}$ and $\phi=\arccos(\frac{A}{\sqrt{A^2+B^2}})$. With these generalized rotation operators in hand, we can apply the same procedure derived in Ref \citenum{Kivlichan2018} and summarized above, but replacing $\hat{R}_{pq}$ by $\hat{R'}_{pq}$, and $\mathbf{r}_{pq}(\theta)$ by $\mathbf{r'}_{pq}(\theta,\phi).$
The values $\theta=\theta_{k}$ and $\phi=\phi_{k}$ for the complex-valued rotation matrix $\mathbf{r'}_{pq}(\theta,\phi)$ that diagonalizes the corresponding complex-valued matrix $\textbf{u}$ (cf. eq.~\ref{eq:qr_diag}) are obtained from a corresponding generalization of the QR-like decomposition that is given explicitly in the ESI.
As a result, the general complex orbital rotation matrix for the Im-uCJ and g-uCJ ansatz can be decomposed into $\binom{N}{2}$  generalized fermionic rotation operators, according to 

\begin{equation}
\hat{U}(u) = \left( \prod_{k=M}^{1} \hat{R'}_{pq}(-\theta_k,\phi_k) \right) \prod_{p=1}^{N} e^{i{\phi'}_p \hat{n}_p}.
\label{eq:qr_unitary_generalized}
\end{equation}

By further restricting the fermionic rotations to adjacent qubits we avoid non-local lengthy JW $\hat{Z}$ Pauli strings and use only local Givens rotations operators. With this realization, the operator $\hat{R'}_{pq}$ is effectively a two-qubit operator, which can be realized with the use of only three CNOT gates.\cite{Vidal2004, Vatan2004, Shende2004} 
%\bwcomm{The values of  $\theta_{k}$ and $\phi_{k}$ for the complex-valued rotation matrix $\mathbf{r'}_{pq}(\theta,\phi)$ that diagonalizes the corresponding complex-valued matrix $\textbf{u}$ (cf. Eq.~\ref{eq:qr_diag} are obtained from a corresponding generalization of the QR-like decomposition that is given explicitly in the ESI.} 

One of the important benefits of using exact exponentiation over Trotter decomposition is the fact that we can effectively treat biradicaloid systems when using the NOQE.\cite{Baek2023} In a typical biradicaloid, pairs of unrestricted Hartree–Fock reference states differ only by permutations of spin orbitals, allowing the same \textbf{K} and \textbf{J} matrices to be reused for each reference simply by permuting the corresponding matrix indices. This reuse significantly reduces the cost of parameter optimization, as only one set of parameters must be optimized. The energy accuracy can then be improved by incorporating additional references with no more classical preprocessing costs, although there is now the  additional quantum processor expense to measure more nondiagonal overlap and Hamiltonian matrix elements. By contrast, in an approximate Trotter decomposition scheme, these parameters 
must generally be re-optimized for each reference state, removing the classical preprocessing efficiency that exact exponentiation affords.

\subsection{Explicit illustration of ansatz construction with \ce{H2} STO-3G}

To illustrate the distinctions between the three uCJ ansätze, we consider an illustrative example of a four spin-orbital, two-electron system,: the \ce{H2} molecule in the STO-3G basis set. Given  spin-orbitals $\phi_{i}^{\sigma}$ ($\sigma = \alpha, \beta$) and using the Jordan-Wigner mapping with the convention that occupied orbitals are listed first, we represent our wavefunction in occupation vector form as follows: $\ket{\phi^\alpha_1\phi^\beta_2\phi^\alpha_3\phi^\beta_4}$, where orbitals $\phi^\alpha_1\phi^\beta_2$ are occupied and $\phi^\alpha_3\phi^\beta_4$ are virtual spin-orbitals in the HF state. The explicit form of the operator $\hat{K}_\mathrm{g-uCJ}$ is:
\begin{equation}
    \label{eq_15}
  \hat{K}_\mathrm{g-uCJ} = K_{13}\hat{a}^{\dag}_{1}{a}_{3}+{K_{24}\hat{a}^{\dag}_{2}{a}_{4}}+{K_{31}\hat{a}^{\dag}_{3}{a}_{1}}+{K_{42}\hat{a}^{\dag}_{4}{a}_{2}}. 
\end{equation}
Since \textbf{K} is anti-Hermitian, we have
\begin{equation}
    \label{eq_16}
  K_{13} = -K_{31}^*, \: K_{24} = -K_{42}^* , 
\end{equation}
which translates to
\begin{equation}
    \begin{aligned}
    \label{eq_17}
  \RE(K_{13}) = -\RE(K_{31}), \: \RE(K_{24}) = -\RE(K_{42}), \\ \IM(K_{13}) = \IM(K_{31}), \: \IM(K_{24}) = \IM(K_{42}) . 
    \end{aligned}
\end{equation}

Now, let us express the $\hat{K}$ operator in terms of Pauli words using the Jordan-Wigner transformation:
\begin{equation}
    \begin{aligned}
    \label{eq_18}
    \hat{K}_\mathrm{g-uCJ} = \frac{1}{4}[(K_{13}+K_{31})(\hat{X}_1\hat{Z}_2\hat{X}_3\hat{I}_4+\hat{Y}_1\hat{Z}_2\hat{Y}_3\hat{I}_4)+\\
    i(K_{31}-K_{13})(\hat{Y}_1\hat{Z}_2\hat{X}_3\hat{I}_4-\hat{X}_1\hat{Z}_2\hat{Y}_3\hat{I}_4)+\\
    (K_{24}+K_{42})(\hat{I}_1\hat{X}_2\hat{Z}_3\hat{X}_4+\hat{I}_1\hat{Y}_2\hat{Z}_3\hat{Y}_4)+\\
    i(K_{42}-K_{24})(\hat{I}_1\hat{Y}_2\hat{Z}_3\hat{X}_4-\hat{I}_1\hat{X}_2\hat{Z}_3\hat{Y}_4)].
    \end{aligned}
\end{equation}
Given the anti-Hermitian nature of the \textbf{K} matrix, the expression above can be rewritten as:
\begin{equation}
    \begin{aligned}
    \label{eq_19}
    \hat{K}_\mathrm{g-uCJ} = \frac{i}{2}[\IM(K_{31})(\hat{X}_1\hat{Z}_2\hat{X}_3\hat{I}_4+\hat{Y}_1\hat{Z}_2\hat{Y}_3\hat{I}_4)+\\
    \RE(K_{31})(\hat{Y}_1\hat{Z}_2\hat{X}_3\hat{I}_4-\hat{X}_1\hat{Z}_2\hat{Y}_3\hat{I}_4)+\\
    \IM(K_{42})(\hat{I}_1\hat{X}_2\hat{Z}_3\hat{X}_4+\hat{I}_1\hat{Y}_2\hat{Z}_3\hat{Y}_4)+\\
    \RE(K_{42})(\hat{I}_1\hat{Y}_2\hat{Z}_3\hat{X}_4-\hat{I}_1\hat{X}_2\hat{Z}_3\hat{Y}_4)].
    \end{aligned}
\end{equation}
Thus, there are four real parameters to optimize for this operator. However, by restricting $\RE(K_{ij})=0$ (Im-uCJ, eq. \ref{eq:Im-uCJ}) or $\IM(K_{ij})=0$ (Re-uCJ, eq. \ref{eq:Re-uCJ}), we reduce the number of parameters to two: 
\begin{equation}
    \begin{aligned}
    \label{eq_20}
    \hat{K}_\mathrm{Im-uCJ} = \frac{i}{2}[\IM(K_{31})(\hat{X}_1\hat{Z}_2\hat{X}_3\hat{I}_4+\hat{Y}_1\hat{Z}_2\hat{Y}_3\hat{I}_4)+\\
    \IM(K_{42})(\hat{I}_1\hat{X}_2\hat{Z}_3\hat{X}_4+\hat{I}_1\hat{Y}_2\hat{Z}_3\hat{Y}_4)],
    \end{aligned}
\end{equation}
\begin{equation}
    \begin{aligned}
    \label{eq_21}
    \hat{K}_\mathrm{Re-uCJ} = \frac{i}{2}[\RE(K_{31})(\hat{Y}_1\hat{Z}_2\hat{X}_3\hat{I}_4-\hat{X}_1\hat{Z}_2\hat{Y}_3\hat{I}_4)+\\
    \RE(K_{42})(\hat{I}_1\hat{Y}_2\hat{Z}_3\hat{X}_4-\hat{I}_1\hat{X}_2\hat{Z}_3\hat{Y}_4)t].
    \end{aligned}
\end{equation}

Since $[\hat{n}_i,\hat{n}_j]=[\hat{a}^{\dag}_{i}\hat{a}_{i},\hat{a}^{\dag}_{j}\hat{a}_{j}]=0$, and given that \textbf{J} is symmetric and purely imaginary, the explicit form of the $\hat{J}$ operator for our two electron, four spin-orbital singlet case is:

\begin{equation}
    \begin{aligned}
    \label{eq_22}
  \hat{J} = 2[J_{12}\hat{a}^{\dag}_{1}\hat{a}_{1}\hat{a}^{\dag}_{2}\hat{a}_{2}+J_{14}\hat{a}^{\dag}_{1}\hat{a}_{1}\hat{a}^{\dag}_{4}\hat{a}_{4}+\\
  J_{23}\hat{a}^{\dag}_{2}\hat{a}_{2}\hat{a}^{\dag}_{3}\hat{a}_{3}+J_{34}\hat{a}^{\dag}_{3}\hat{a}_{3}\hat{a}^{\dag}_{4}\hat{a}_{4}]
    \end{aligned}
\end{equation}
Using the JW transformation, we can show that:
\begin{equation}
    \begin{aligned}
    \label{eq_23}
\hat{J} =\frac{1}{2}[(J_{12}+J_{14}+J_{23}+J_{34})\hat{I}_1\hat{I}_2\hat{I}_3\hat{I}_4
    -(J_{12}+J_{14})\hat{Z}_1\hat{I}_2\hat{I}_3\hat{I}_4\\
    -(J_{12}+J_{23})\hat{I}_1\hat{Z}_2\hat{I}_3\hat{I}_4
    -(J_{23}+J_{34})\hat{I}_1\hat{I}_2\hat{Z}_3\hat{I}_4\\
    -(J_{34}+J_{14})\hat{I}_1\hat{I}_2\hat{I}_3\hat{Z}_4
    +J_{12}\hat{Z}_1\hat{Z}_2\hat{I}_3\hat{I}_4+J_{14}\hat{Z}_1\hat{I}_2\hat{I}_3\hat{Z}_4\\+J_{23}\hat{I}_1\hat{Z}_2\hat{Z}_3\hat{I}_4+J_{34}\hat{I}_1\hat{I}_2\hat{Z}_3\hat{Z}_4].
    \end{aligned}
\end{equation}
This yields four variational parameters to optimize. Note that we assumed $\hat{n}_1\hat{n}_3\ket{\Psi}=\hat{n}_2\hat{n}_4\ket{\Psi}=0$, given the singlet multiplicity of the wavefunction for ground state H$_2$. In the general case, without this assumption, there would be six variational parameters to optimize for \textbf{J}.

\subsection{Technical details}

We have examined several chemical systems, \ce{H2}, \ce{H3+}, \ce{Be_2}, \ce{C2H4}, \ce{C2H6} and \ce{C6H6}, to evaluate and compare the performance of the VQE algorithm with each of the three uCJ ansätze. The RHF solutions were generated using the PySCF software package \cite{sun2020recent}. The exponentiation of the $\hat{K}$ operator in the Re-, Im-, and g-uCJ circuits was implemented exactly through the Givens rotation as described in section 2.2 above. For the \ce{C2H6}, \ce{Be2}, \ce{C2H4}, and \ce{C6H6} molecules, the CASCI formalism was used with active spaces of (2e,2o),(2e,2o),(4e,4o), and (4e,4o),  respectively (we denote the number of spatial orbitals o here, according to quantum chemistry convention, and use $N$ to denote the number of spin orbitals, according to quantum computing convention). 
The corresponding active space orbitals are illustrated in Figure S2. Optimization of the \textbf{K} and \textbf{J} matrices was carried out classically with an implementation of the SLSQP minimizer,\cite{Kraft1988} with a sparse-vector representation of quantum states. For larger active spaces (greater than or equal to 8 qubits), optimization of the \textbf{K} and \textbf{J} matrices was first performed in a reduced space under the perfect pairing assumption (only one bonding and its corresponding antibonding orbital are entangled at a time through the \textbf{K} and \textbf{J} operators). 
The orbital connectivity was then gradually extended toward the fully connected case, using the optimized coefficients from the previous step as the initial guess for each subsequent stage. The simulator implementation was written locally, and can be accessed through GitHub.\cite{TkachenkoGitHub} All numerical simulations presented in this work were carried out in the noiseless setting. The UCCSD simulations were performed with the Qiskit-nature package.\cite{QiskitNature2023,qiskit_nature_webpage} For the single Trotter-step decomposed ansätze (UCCSD, Re-uCJ, Im-uCJ, and g-uCJ), the number of CNOT gates was computed by first performing the Jordan-Wigner transformation of the $\hat{T}_1$, $\hat{T}_2$, $\hat{J}$ and $\hat{K}$ operators, followed by simplifying the resulting Pauli word sums through term cancellation, and finally representing each remaining exponentiated Pauli word by efficient quantum circuits as described in Ref. \citenum{PhysRevA.102.062612}.
For exact implementations of the uCJ ansätze, the exponential $e^{\hat{J}}$ was similarly decomposed by obtaining its JW representation, recognizing that in this case all Pauli words contain only $Z$ and $I$ gates and thus pairwise commute, allowing the exact decomposition into a product of exponentials of individual Pauli words.
The $e^{\hat{K}}$ and $e^{-\hat{K}}$ terms were implemented by applying $2\binom{N/2}{2}$ generalized orbital rotation operators $R'_{pq}(\theta,\phi)$ as described in Section 2.2, where $N$ represents the total number of spin orbitals, and division by two accounts for spin-orbital rotations only within the same spin space.

%\bwcomm{BW: might be useful to add a bit more detail here, I did not see any documentation on the Qistki-nature page} The CNOT gate was assumed as a native two-qubit gate within this work.

\section{Results for molecular benchmarks}
\subsection{Circuit depth analysis}
We begin our discussion with an analysis of circuit depth for the investigated ansätze. As shown in Table \ref{tbl_1}, the exact implementations of the three uCJ ansatz variants result in a substantially smaller number of native two-qubit gates (CNOTs) compared to UCCSD circuits. As expected, this difference becomes more pronounced for larger systems. For example, for the 12-qubit \ce{H2} system in the 6-311G basis, the exact implementations of all uCJ variants require approximately six times fewer two-qubit gates than the single-step Trotter-Suzuki decomposition of UCCSD (192 vs. 202 two-qubit gates). We note that this dominance is not observed when using single-step Trotter-Suzuki decompositions of the uCJ ansätze. In fact, Table~\ref{tbl_1} shows that in this situation, the g-uCJ variant can in some cases perform worse than UCCSD. Therefore, we recommend using exact implementations of the uCJ ansätze whenever possible.

We now compare the performance of the restricted versus general uCJ ansätze. For single-step Trotterized circuits, the restricted Re-uCJ and Im-uCJ variants of uCJ employ approximately two-thirds the number of two-qubit gates compared to the generalized g-uCJ ansätz. This behavior arises from the two-fold increase in the number of Pauli terms in the definition of the $\hat{K}$ operator for g-uCJ (see Eqs.~\ref{eq_19} - \ref{eq_21}).
In contrast, for the exact implementations  of Im-uCJ and g-uCJ introduced in this work that use the generalized Givens rotations, the CNOT cost is similar across all three uCJ variants in their exact form (see Eqs.~\ref{eq:qr_diag} - \ref{eq:givens_rotation2}), although the number of variational parameters differs between restricted and generalized variants. 
%\bwcomm{BW: ok the number of variational parameters is obviously different, but why is the number of CNOT gates the same?}

As we demonstrate in the performance analysis section below, the increased number of variational parameters in g-uCJ is offset by its greater flexibility and representability. In all of the considered cases, the g-uCJ ansatz recovers a larger portion of the correlation energy than its restricted counterparts and often yields results within chemical accuracy. If the number of variational parameters is a critical constraint, we recommend using the restricted variants (Re-uCJ or Im-uCJ), with the remaining correlation energy recovered using, for example, the NOQE algorithm.\cite{Baek2023} 

\begin{table*}[h!]
    \centering
    \small
    \caption{Comparison of the number of two-qubit gates for the systems considered in this manuscript. $N_{2q}$ stands for the number of native two-qubit gates (CNOTs) in the circuit, "single T-step" refers to the Trotter-Suzuki decomposition with one step.}
    \label{tbl_1}
    \begin{tabular}{|c|c|c|c|c|c|c|c|c|c|}
        \hline
        \multirow{2}{*}{System} & \multirow{2}{*}{$N_{elec}$} & \multirow{2}{*}{$N_{qubit}$} & UCCSD $N_{2q}$ & 
        \multicolumn{2}{c|}{g-uCJ $N_{2q}$} & 
        \multicolumn{2}{c|}{Re-uCJ $N_{2q}$} & 
        \multicolumn{2}{c|}{Im-uCJ $N_{2q}$} \\
        \cline{4-10}
        & & & single T-step & single T-step & exact & single T-step & exact & single T-step & exact \\
        \hline
        $\ce{H2}$ (STO-3G)            & 2 & 4  & 42   & 46  & 20 & 36  & 20 & 34  & 20 \\
        $\ce{Be2}$ (STO-3G, (2e,2o))  & 2 & 4  & 42   & 46  & 20 & 36  & 20 & 34  & 20 \\
        $\ce{C2H6}$ (STO-3G, (2e,2o)) & 2 & 4  & 42   & 46  & 20 & 36  & 20 & 34  & 20 \\
        $\ce{H3+}$ (STO-3G)           & 2 & 6  & 166 & 166 & 54 & 120 & 54 & 126 & 54 \\
        $\ce{H2}$ (6-31G)             & 2 & 8  & 404 & 376 & 104 & 264 & 104 & 270 & 104 \\
        $\ce{H4}$ (STO-3G)            & 4 & 8  & 753 & 400 & 128 & 288 & 128 & 294 & 128 \\
        $\ce{C2H4}$ (STO-3G, (4e,4o)) & 4 & 8  & 753 & 400 & 128 & 288 & 128 & 294 & 128 \\
        $\ce{C6H6}$ (STO-3G, (4e,4o)) & 4 & 8  & 753 & 400 & 128 & 288 & 128 & 294 & 128 \\
        $\ce{H2}$ (6-311G)            & 2 & 12 & 1202 & 1319 & 192 & 894 & 192 & 902 & 192 \\
        \hline
    \end{tabular}
    \label{tab:system-details}
\end{table*}

\subsection{Performance analysis}

We begin our performance analysis with the H$_2$ molecule in the STO-3G basis set. Bond dissociation curves were calculated to evaluate the performance of the three uCJ ansätze across different correlation regimes. 
%\bwcomm{BW: Nikolay, the next 3 sentences were confusing to me, so I reordered and reworded them, making some changes after looking at the ESI carefully. Please check my changes! The original is still there, commented out.}
%In this system, the UHF solution transitions into a symmetry-broken wavefunction beyond a bond distance of 1.2 Å (Coulson-Fischer point). While the RHF reference was considered for all molecules, a comparison between results for H$_2$ obtained from UHF and RHF reference states is provided in the Supplementary Information (Figure S1), which shows that RHF performs better in the single-reference VQE regime inside the Coulson-Fischer point). 
We considered an RHF reference for all molecules, but for H$_2$ we also compared results obtained from UHF reference states. 
%As expected, we found that single-reference ansätze built on RHF references perform better inside the Coulson-Fischer point, which is intrinsically a single-reference regime. 
Beyond the Coulson-Fischer point, a single-reference ansatz based on UHF becomes a symmetry-broken wavefunction and the resulting energies show greater error than the ansätze built on RHF references (see Figure S1 in the ESI). We also see that for all except very small internuclear distances, the energies obtained with single-reference Im-uCJ are systematically lower than those obtained with single-reference Re-uCJ.
Notably, the g-uCJ ansatz exhibits sufficient flexibility to exactly reproduce the FCI energies for H$_2$ in STO-3G, a trend that holds across all the two-electron problems examined in this work.

\begin{figure*}[h!]
\centering
  \includegraphics[height=16cm]{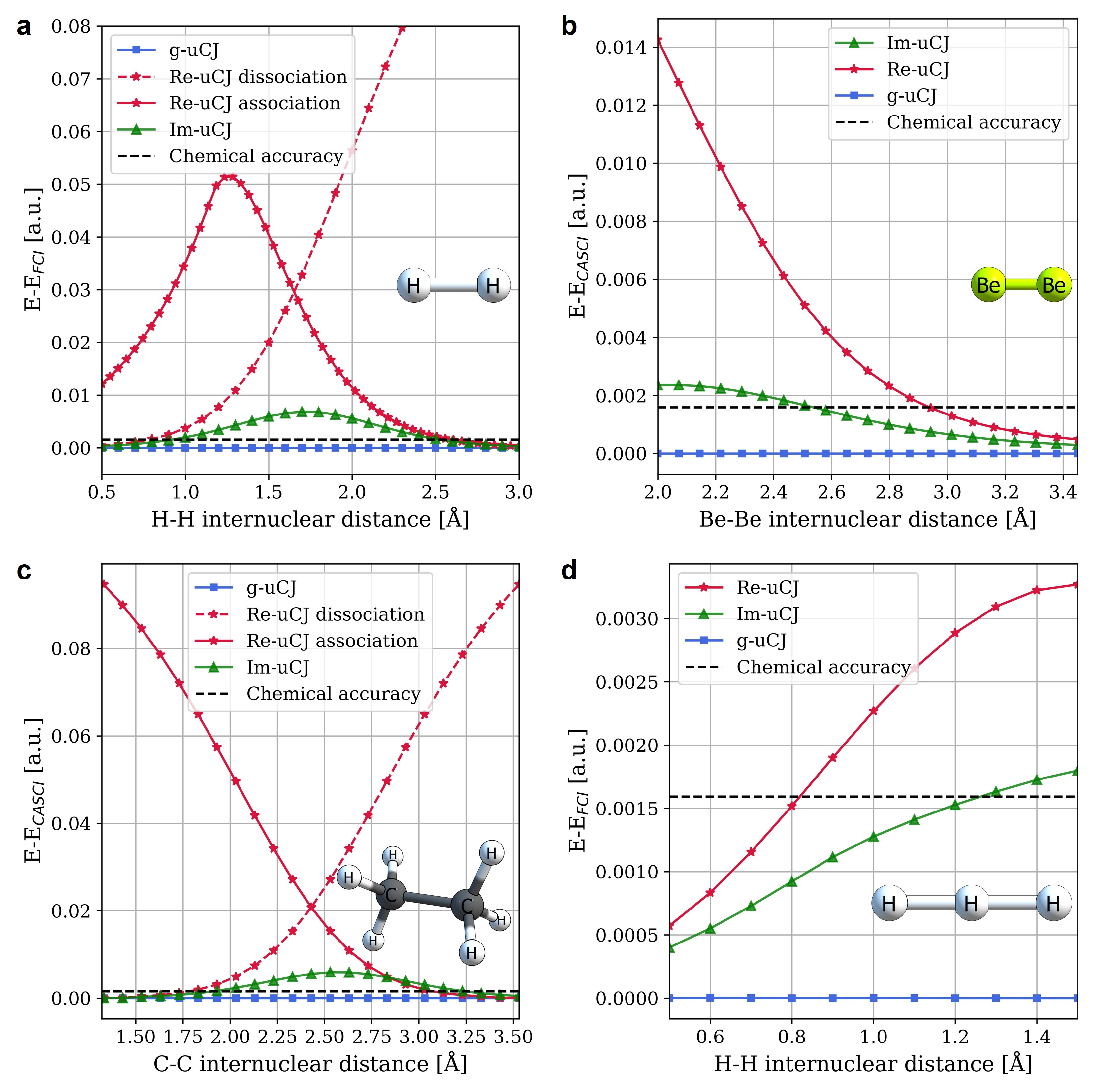}
  \caption{Errors in energies for variational single reference calculations with g-uCJ, Re-uCJ and im-uCJ ans\"atze relative to FCI or CASSI energies for (a) \ce{H2}, (b) \ce{Be2} (2e,2o), (c) \ce{C2H6} (2e,2o), and (d) linear \ce{H3+}. The  STO-3G basis set was used for all molecules. The horizontal black dashed line corresponds to chemical accuracy (1 kcal/mol). The RHF state was used as a reference state for all systems. The curves denoted by 'association' or 'dissociation' differ in how the initial guesses for the $\mathbf{K}$ and $\mathbf{J}$ matrices were obtained. For the association curves, the optimized parameters from the prior step were used as the initial guess at the next step as the bond distance was gradually decreased. In contrast, for the dissociation curves, the bond distance was gradually increased.}
  \label{fig1}
\end{figure*}

\begin{figure}[h!]
\centering
  \includegraphics[height=8cm]{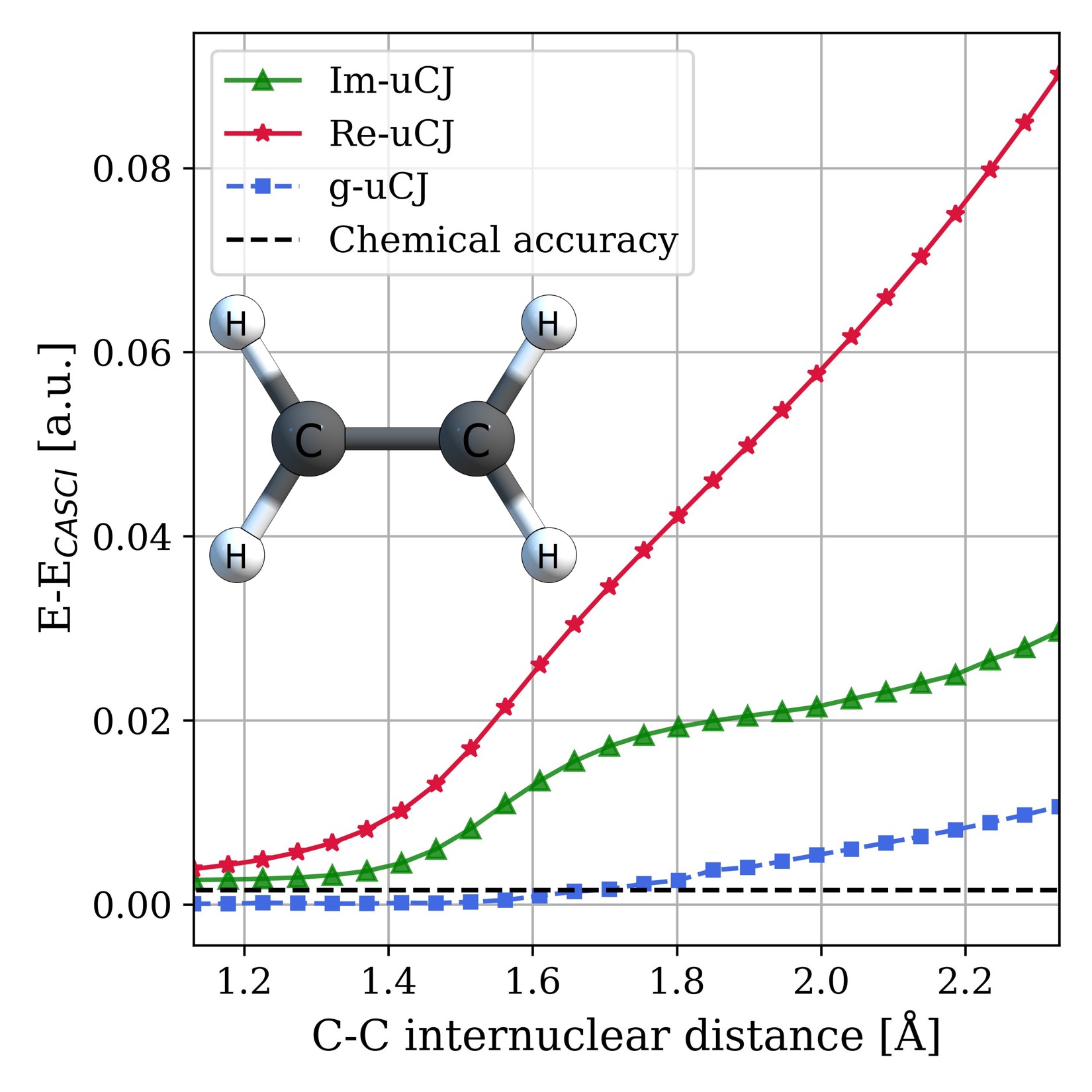}
  \caption{Errors in energies for g-uCJ, Re-uCJ and im-uCJ relative to CASSI energies for \ce{C2H4} (4e,4o) at STO-3G basis set. The horizontal black dashed line corresponds to chemical accuracy (1 kcal/mol). The RHF state was used as a reference state.}
  \label{fig_C2H4}
\end{figure}

As illustrated for \ce{H2}, \ce{Be_2} (2e,2o), \ce{C2H6} (2,2o), and linear \ce{H_3^+} in Figure \ref{fig1}, the performance of the Im-uCJ and Re-uCJ ans\"atze show significant differences. 
%\bwcomm{BW: are all of these and remaining calculations using RHF ansatze? It would be useful to state this again explicitly here and also in the captions to Figs 1-3. Also, can you explain your notation of 'dissociation' and 'association'? NVT: added to the caption} 
Despite requiring the same number of two-qubit gates, Im-uCJ consistently yields more accurate results, whereas Re-uCJ sometimes converges to metastable local minima and struggles with optimization. This behavior is clearly illustrated in Figures \ref{fig1}a and \ref{fig1}b. Here the Re-uCJ curves denoted as 'dissociation' were obtained by gradually increasing the bond distance and using the optimized coefficients from each prior step as an initial guess. This approach accumulates errors as the bond breaks, as is also seen and well-known for RHF ans\"atze. The energy can be reduced by variational optimisation of the Re-uCJ parameters, but the results remain inferior to Im-uCJ (except for the \ce{C2H6} case with $R_{CC}> 2.9$ \AA, where Re-uCJ performs slightly better than Im-uCJ), and this variational local minimum becomes inferior at shorter bond distances ($R_{CC}< 1.6$ \AA\space for \ce{H2}, and $R_{CC}< 2.4$ \AA\space for \ce{C2H6}).
The curves denoted as 'association' were generated by gradually decreasing the bond distance, with each calculation initialized using the optimized coefficients from the previous step.
 We note that random initialization of the Re-uCJ parameters along the dissociation curve leads to erratic optimization behavior due to jumping between the dissociation and association curves, 
 %(where analogously to dissociation curve, to generate the association plot, the bond distance was gradually decreased, with each calculation initialized using the optimized coefficients from the previous step),  
 and disrupting smoothness of the energy as a result. 
In contrast, Im-uCJ reliably converges to a single lower energy solution regardless of the initial guess procedure, demonstrating greater stability. 
%\bwcomm{BW: was this done with using optimized coefficients from each prior step as initial guess?  pls specify which method was used in this case}

Similar results are observed for the other molecular systems shown in Figures \ref{fig1}b-c, although there are some specific %aspects 
features worth commenting on. In particular, we see that \ce{Be2} exhibits no binding in calculations using an RHF reference, while in fact it is known from experiments to have a weak bond, with $R_e = 2.45$ \AA. Part of its strong correlation effect can be captured with a (2e,2o) active space,
allowing $(\sigma^*_{2s})^2 \rightarrow (\sigma_{2p_z})^2$ excitations, as illustrated in Figure S2b. The electron correlations are stronger at $R_e$ than at dissociation, as reflected in the sharp rise in the Re-uCJ error at short distances in Figure \ref{fig1}b. Similar to the dissociation of \ce{H2}, for Be$_2$ we see that Im-uCJ again provides dramatic improvement over Re-uCJ, while g-uCJ attains exactness at all distances. 

While exactness for two-electron systems is a desirable property and %is satisfied 
can be achieved for H$_2$ by optimization of both g-uCJ and UCCSD ans\"atze, realistic applications typically involve larger numbers of electrons in the active space. 
To assess the performance of the uCJ variants in such settings, we examined the case of a 4-electron, 4-orbital system corresponding to the double bond breaking in \ce{C2H4}. 
The results are shown in Figure~\ref{fig_C2H4}. 
As can be seen here, the same general trend persists, namely that g-uCJ performs best, while Re-uCJ yields the least accurate results. 
Notably, g-uCJ maintains chemical accuracy up to a C–C bond distance of 1.7 \AA. 
At larger separations, its performance degrades due to the inability to fully describe the dissociation into two $^3$\ce{CH2} radicals, a process that involves quadruple excitations, which requires going beyond the $k=1$ level of a uCJ ansatz. 
%\bwcomm{BW: would this not also be helped by using UHF references instead of RHF? NVT: upcomming calcs} 
Thus, in a situation where bonds with double or higher order bond character are broken at larger distances, it is recommended to go beyond the $k=1$ level of uCJ ansatz. 
%\bwcomm{BW: similarly here, would it not be useful to consider uCJ ansatze in this situation? and to go to NOQE...??}

For larger systems, single-point energy calculations were performed using the g-uCJ, Re-uCJ, and Im-uCJ ansätze. These are summarized in Table~\ref{tbl_2}, where the previously observed trends are maintained. Thus, Re-uCJ generally recovers the smallest fraction of the correlation energy, Im-uCJ performs better, and of the three ans\"atze, g-uCJ captures the largest fraction of the correlation energy. A comparison with UCCSD is also provided in Table~\ref {tbl_2}. As shown there, g-uCJ achieves accuracy comparable to UCCSD while requiring significantly fewer two-qubit gates. Since symmetry breaking can contribute to the correlation energy in some systems, we also report the percentage of correlation energy captured by a single broken-symmetry UHF reference state, 
%alone, 
without dressing by a cluster Jastrow operator. As shown in line 2 of Table 2 for the highly correlated \ce{H4} 
%square 
molecule in a square geometry, using a bare UHF reference state captures approximately $\approx87\%$ of the electronic correlation energy, while applying the g-uCJ ansatz to the RHF reference improves the accuracy, recovering around $\approx95\%$ of the correlation energy and bringing the result significantly closer to the FCI energy. To further improve upon the accuracy achieved with g-uCJ, algorithms such as NOQE and, in the long term, QPE can be employed.

\begin{table}[h!]
    \caption{Percentage of the total correlation energy (relative to FCI or CASCI solutions) that is recovered with g-uCJ, Re-uCJ, and im-uCJ ans\"atze for three different molecular systems.}
\begin{adjustwidth}{-1cm}{-1cm}
    \centering
    \small
    \label{tbl_2}
    \begin{tabular}{|c|c|c|c|c|c|}
        \hline
        System & UCCSD & g-uCJ & Re-uCJ & im-uCJ & UHF \\
        \hline
        H$_2$ (6-31G, $R_{HH}=1.2Å$)                & 100 & 100 & 82.88 & 99.96 & 0.06 \\
        H$_4$ (STO-3G, $R_{HH}=1.1Å$)               & 92.84 & 94.56   & 89.76   & 92.01 & 87.02   \\
        C$_6$H$_6$ (STO-3G, (4e,4o)) & 100 & 92.26   & 59.74   & 83.97  & N/A  \\
        \hline
    \end{tabular}
    \label{tab:system-details}
\end{adjustwidth}
\end{table}

\subsection{Configuration state function analysis}

\begin{figure*}[h!]
\centering
  \includegraphics[height=16cm]{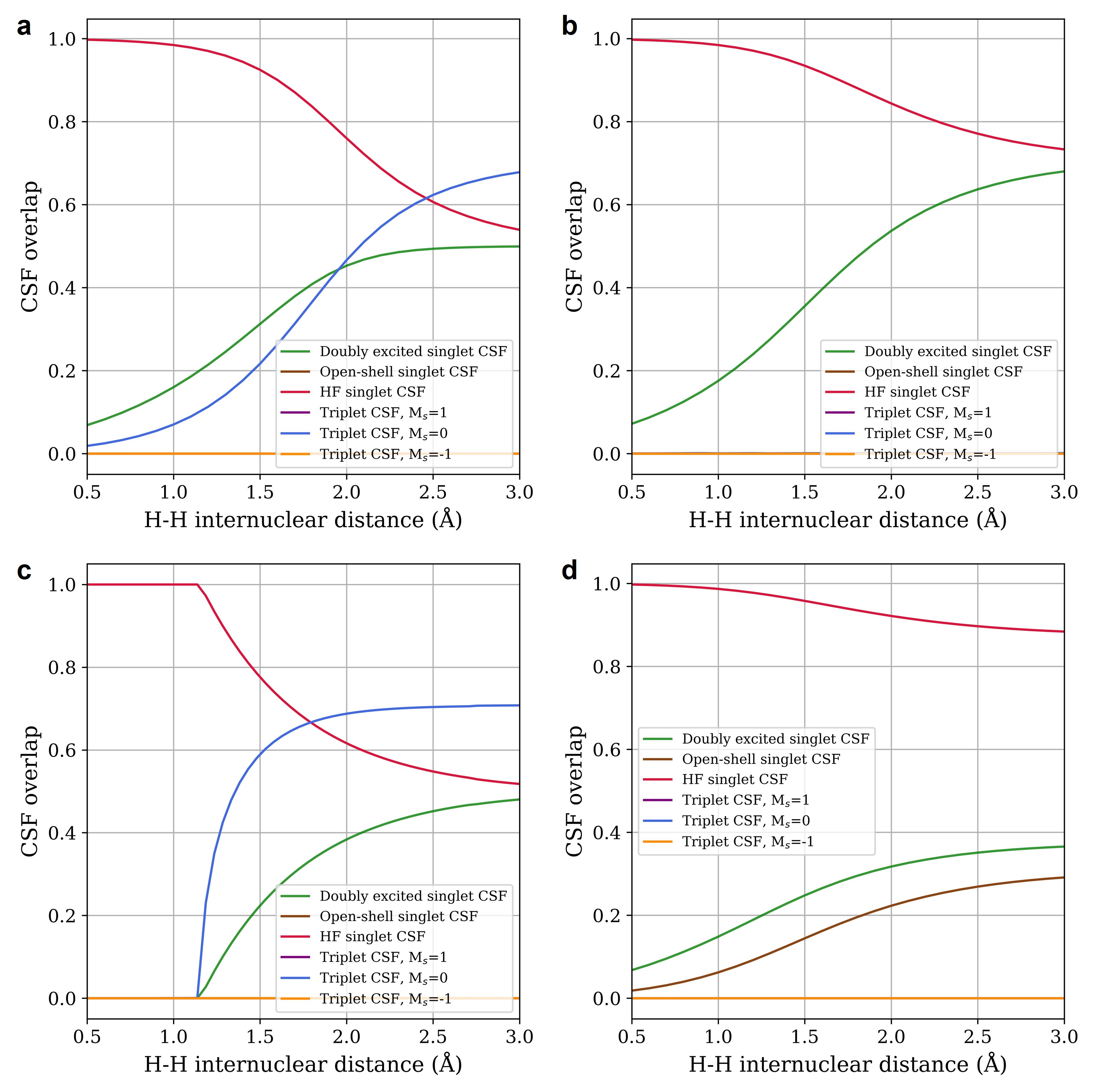}
  \caption{Overlaps of the configuration state functions (CSF) with the wavefunction obtained after application of (a) im-uCJ, (b) g-uCJ, (c) Re-uCJ association, and (d) Re-uCJ dissociation ansätze for the \ce{H2} molecule in the STO-3G basis set. The RHF state was used as a reference state for all systems. The association and dissociation curves for a given ansatz differ in how the initial guesses for the $\mathbf{K}$ and $\mathbf{J}$ matrices were obtained. For association (dissociation) curves, the optimized parameters from the prior step was used as the initial guess at the next step as the bond distance was gradually decreased (increased)
  %\bwcomm{In contrast}, for \bwcomm{dissociation curves}, the bond distance was gradually increased \bwcomm{
  (see text).}
  \label{fig2}
\end{figure*}

To better understand why Im-uCJ yields more accurate energies than Re-uCJ, we performed a configuration state function (CSF) composition analysis of the corresponding wavefunctions for \ce{H2}. The results are summarized in Figure \ref{fig2}. The exact FCI ground state wavefunction should include only two singlet CSFs: the Hartree-Fock state and the doubly excited singlet state, both of grade (g) symmetry. This expected behavior is observed for the g-uCJ ansatz in Figure \ref{fig2}b, where only these two CSFs contribute to the wavefunction, leading to a solution that is correct in both energy, spatial, and spin symmetry.

Interestingly, we find that the Im-uCJ ansatz exhibits a more flexible structure. As shown in Figure \ref{fig2}a, in addition to the HF and doubly excited CSFs, the wavefunction also includes contributions from the triplet $M_s=0$ CSF. 
As is familiar in UHF, this additional flexibility lowers the energy and brings the result closer to the exact FCI energy, albeit at the cost of mixing different spin states. 
In contrast to UHF, the contribution from the triplet CSF decreases smoothly at short distances, but is still non-zero at bond distances 0.25 \AA \ shorter than $R_e$: there is no analog of a Coulson-Fischer point, i.e., no collapse to RHF solution. 
%As a result %Notably, due to the restricted form of the $\hat{K}$ operator, the $M_s=0$ symmetry is preserved, though 
Consequently, the value of $\langle S^2 \rangle$ deviates from zero (a pure singlet) all along the potential energy curve.

The Re-uCJ ansatz displays an even more intriguing behavior, showing very distinct wavefunction compositions for the two local minima deriving from the equilibrium (dissociation ansatz) and separated atom regimes (association ansatz).
%\bwcomm{BW: not clear to me what dissociation and association ansatze are, and why dissociation is in equilibrium regime, association in separated atom regime. If you define these explicitly earlier in the text all will be clear.} 
In the dissociation limit (Figure \ref{fig2}d), the wavefunction deriving from equilibrium retains significant overlap with the HF state, increasing the overall energy error at large distances. Additionally, we observe contributions from the open-shell singlet CSF, indicating that while the $\langle S^2 \rangle$ and $\langle M_s \rangle$ are correct for a pure singlet state, the Re-uCJ ansatz exhibits spatial symmetry breaking (the open-shell singlet has ungerade (u) symmetry). 

In contrast, the Re-uCJ local minimum derived from the separate atom regime (Figure \ref{fig2}c) behaves quite differently. Initially, in the dissociation regime, the wavefunction contains a mixture of triplet $M_s=0$, doubly excited singlet, and HF singlet CSFs, but as the bond distance decreases, the wavefunction collapses into the RHF solution, rationalizing the energy curve behavior observed in Figure \ref{fig1}a. Therefore this solution exhibits the analog of a Coulson-Fischer point, and appears to approach 50\% triplet character at dissociation, resembling a UHF reference state. 
Interestingly, its spin contamination exceeds that of the Im-uCJ solution for all distances beyond about 1.2 \AA~as can be seen from the larger contribution form the triplet $M_s=0$ CSF. 
%\bwcomm{BW: need to document this if we discuss it, e.g., with a table or in the text with numbers here}

\subsection{Two-qubit gate costs for 
%bare 
unencoded qubit calculations}
We estimate here the scaling of the number of gates required for the circuits generating the uCJ ansätze (exact implementation without Trotter decomposition) with bare qubits, i.e., without using any encoding into a quantum error code.  In all gate estimates, we assume an all-to-all qubit connectivity in the quantum device, compatible with trapped ion architectures. We assume compilation of the circuits into CNOT gates and arbitrary single qubit rotations $R_z$, where the latter can be further decomposed into of order $1.15\cdot log_2(1/\epsilon_{syn})+9.2$ T gates, with arbitrary synthesis error $\epsilon_{syn}$.\cite{Bocharov2015}

The exponential of the $\hat{K}$ operator ($e^{\hat{K}}$) can be represented as a series of Givens rotations~\cite{Kivlichan2018}, which requires $\binom{N}{2}$ operations, where $N$ is the number of spin-orbitals. Given that our generalized Givens rotation is essentially a two-qubit gate, it can be in general represented with 3 CNOT gates.\cite{Vidal2004, Vatan2004, Shende2004}
The exponential of the $\hat{J}$ operator ($e^{\hat{J}}$), which consists of paired number operator rotations of the form $e^{-i\theta\hat{n}_i\hat{n}_j}$, requires two CNOT gates and one $R_z$ gate per term. Noting that there are $\binom{N}{2}$ distinct number operator pair products (without diagonal terms), one can then estimate the maximum number of CNOT gates as:

\begin{equation}
N_{\text{CNOT}}^{e^{\hat{K}}} = N_{\text{CNOT}}^{e^{-\hat{K}}} = \binom{N}{2} \times 3 \tag{24}
\end{equation}

\begin{equation}
N_{\text{CNOT}}^{e^{\hat{J}}} = \binom{N}{2} \times 2 \tag{25}
\end{equation}

This results in total counts of at most 
\begin{equation}
N_{\text{CNOT}}^{\text{total}} = N_{\text{CNOT}}^{e^{\hat{K}}} + N_{\text{CNOT}}^{e^{\hat{J}}} + N_{\text{CNOT}}^{e^{-\hat{K}}} = \binom{N}{2} \times 5 \tag{27}
\end{equation}
\noindent for the circuits to generate individual uCJ ansatz states.

\section{Conclusions}
In this work, we introduced two new variants of the $k$-fold unitary cluster Jastrow ansatz that are suitable for variational, subspace expansion, or non-orthogonal quantum eigensolver approaches. 
These are the imaginary cluster Jastrow, Im-uCJ, and the general cluster Jastrow, g-uCJ. We evaluated their performance here within a variational framework using single reference states, and choosing the simplest $k=1$ model.  Specifically, we showed that similarly to Re-uCJ it is possible to construct an exact exponentiation procedure for the Im- and g-uCJ ansätze that incorporates generalized orbital rotations, thereby completely avoiding the need for Trotter decomposition.

Our $k=1$ results show that the restricted Im-uCJ ansatz has the best performance, offering a compelling balance between accuracy and computational efficiency, yielding shallow circuits, a moderate number of parameters for optimization, and superior performance compared to the previously proposed Re-uCJ ansatz, which often struggles with local minima and convergence issues. The generalized g-uCJ ansatz achieves near-exact accuracy, even reproducing FCI energies for small systems, however it does so at the cost of requiring a larger number of parameters. 

A configuration state function (CSF) analysis provided deeper insight into the behavior of these uCJ ansätze. We observed that Im-uCJ lowers the energy at the cost of mixing spin states, while g-uCJ maintains the correct spin symmetry and achieves highly accurate results, albeit with more parameters. On the other hand, Re-uCJ exhibits a strong dependence on initialization, with distinct behaviors in association and dissociation

Given the strong performance of Im-uCJ and g-uCJ, our future objectives are to apply the proposed uCJ circuits within the NOQE algorithm to explore their potential for multi-reference quantum eigensolvers.

In particular, for biradical systems with two broken-symmetry UHF reference states that differ only by a permutation of $\alpha$ and $\beta$ electrons, the non-orthogonal quantum eigensolver algorithm can improve energy estimates without requiring reoptimization of uCJ parameters. %\bwcomm{BW: cite next paper??} 
In such cases, only a single parameter set needs to be optimized for Im- or g-uCJ, with the second set readily obtained via index permutation in the \textbf{K} and \textbf{J} matrices. This analysis implies that uCJ ansätze, particularly Im- and g-uCJ, are strong candidates for near-term quantum simulations, offering viable and compact alternatives to more resource-intensive methods like UCCSD.

\section*{Author contributions}
The article was written from the contribution of all authors.

\section*{Conflicts of interest}
There are no conflicts to declare.

\section*{Data availability}

The data supporting this article have been included as part of the Supplementary Information. Any related additional data are available upon request from the authors.

\section*{Acknowledgements}

This work was partially supported by a joint development agreement between UC Berkeley and Dow, by the National Science Foundation (NSF) Quantum Leap Challenge Institutes (QLCI) program through Grant No. OMA-2016245,
and by the U.S. Department of Energy, Office of Science, Office of Advanced Scientific Computing Research under Award Number DE-SC0025526.
%National Quantum Information Science Research Centers, Quantum Systems Accelerator. 
WMB is supported by the National Science Foundation Graduate Research Fellowship Program under Grant No. DGE 2146752. Any opinions, findings, and conclusions or recommendations expressed in this material are those of the authors and do not necessarily reflect the views of the National Science Foundation.

%\hang{\begin{table*}[h]
%\centering
%\caption{Summary of performance of UCCSD, Re-uCJ, Im-uCJ, and g-uCJ ansätze.}
%\begin{tabular}{lcccc}
%\hline
%Feature & UCCSD & Re-uCJ & Im-uCJ & g-uCJ \\ 
%\hline
%Orbital rotation coefficients & Real & Real & Imaginary & Complex \\
%Number of variational parameters & High & Low & Low & Moderate \\ 
%Correlation energy recovery & Very High (92--100\%) & Moderate (0--90\%) & High (83--100\%) & Very High (92--100\%) \\
%Two-qubit gate cost & High & Low & Low & Low \\ 
%Convergence stability & Good & Poor (multiple minima) & Good & Good \\ 
%\hline
%\end{tabular}
%\label{tab:ucj_comparison}
%\end{table*}}


\balance

\renewcommand\refname{References}

%%%REFERENCES%%%
\bibliography{rsc} %You need to replace "rsc" on this line with the name of your .bib file
\bibliographystyle{rsc} %the RSC's .bst file



\section*{Supplementary material}

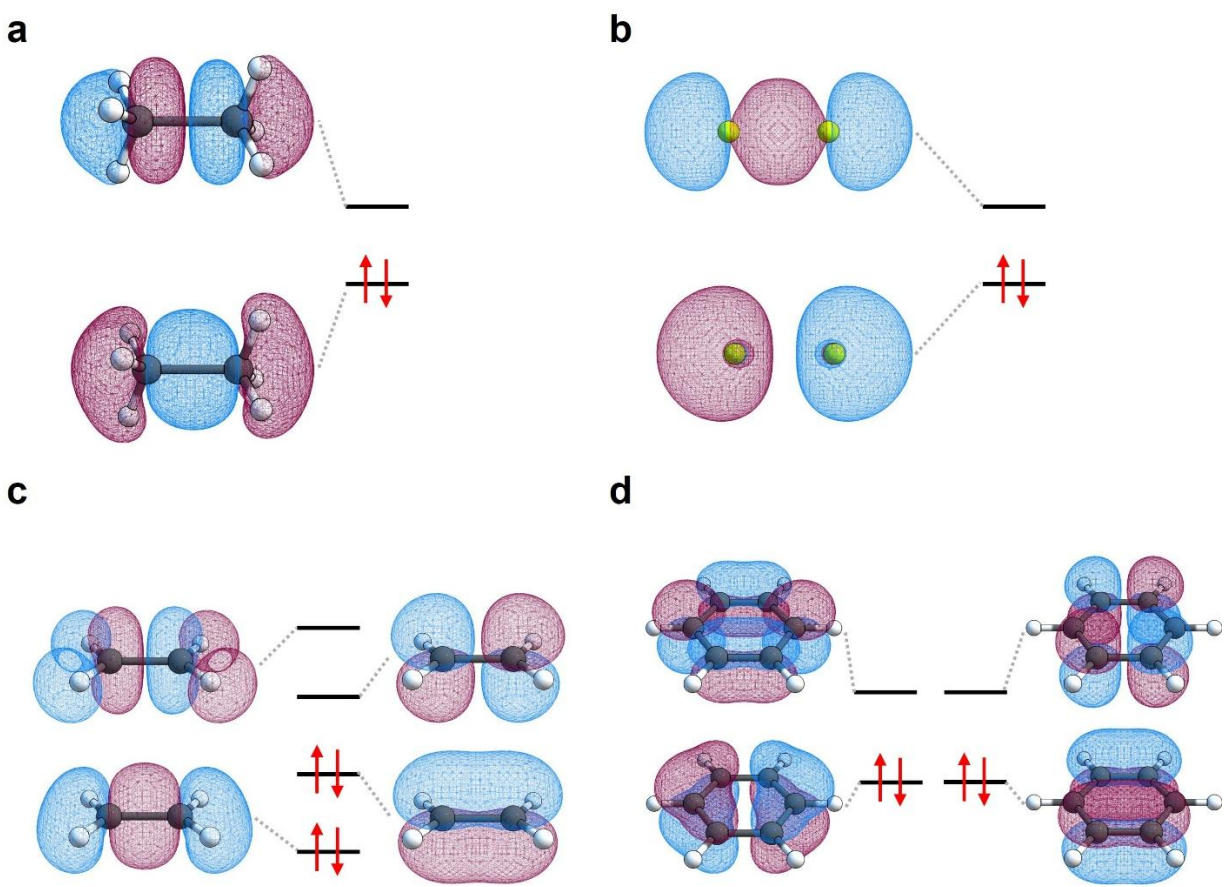

**Figure S2.** Illustration of active space orbitals used for (a) $C_2H_6$ (2e,2o), (b) $Be_2$ (2e,2o), (c) $C_2H_4$ (4e, 4o), and (d) $C_6H_6$ (4e, 4o) molecular systems.

**Discussion of the performance with RHF vs UHF reference states.**

Using the $H_2$ molecule in the STO-3G basis set as an illustrative example (Figure S1), we clearly observe that beyond the Coulson–Fischer point at internuclear distance 1.153 Å, the RHF reference becomes a more favorable choice than UHF when only a single reference is employed. For the Re-uCJ ansatz, the only potential advantage of using a single UHF reference lies in the stability of the optimization. Unlike the RHF-based Re-uCJ, the UHF-based version did not exhibit erratic behavior or multiple local minima during the optimization process. However, this benefit does not extend to the Im-uCJ ansatz, for which such instabilities were not observed with the RHF reference. Therefore, UHF offers no clear advantage in that context. Despite this, the broken-symmetry nature of UHF nevertheless allows for the application of quantum subspace diagonalization (QSD) methods or related methods such as the NOQE algorithm (ref. 55 of the main text). When used in conjunction with Im-uCJ, this can lead to further improvements in performance, providing justification for the use of UHF references in this situation. Finally, for the g-uCJ ansatz, both RHF and UHF references yield

exact FCI results, demonstrating the robustness of the general uCJ ansatz with respect to the choice of underlying reference state.

## Discussion of the diagonalization of matrix $u$.

Since the matrix $u$ is generally complex in the Im-uCJ and g-uCJ cases, its elements can be written as

$$u_{ij} = r_{ij}e^{(i\varphi_{ij})} \; ; \; r_{ij} \in \mathbb{R}; \; \varphi_{ij} \in [0,2\pi).$$

Following the description provided in Ref. 68, when the complex Givens rotation matrix $r'_{pq}(\theta,\phi)$, given in Eq. (16) of the main text left-multiplies the N by N unitary matrix $u$, the resulting product (matrix $u'$) is also unitary. The modified entries of new matrix $u'$ can be written as:

$$u'_{ij} = \begin{cases} \cos(\theta)\, r_{pj}e^{(i\varphi_{pj})} + \sin(\theta)\, r_{qj}e^{(i(\varphi_{qj}+\phi))} & if \; i = p \\ \sin(\theta)\, r_{pj}e^{(i(\pi+\varphi_{pj}-\phi))} + \cos(\theta)\, r_{qj}e^{(i\varphi_{qj})} & if \; i = q \\ r_{ij}e^{(i\varphi_{ij})} & if \; i \neq p,q \,. \end{cases}$$

To diagonalize $u$ we will use the angles $\theta = \theta_k$ and $\phi = \phi_k$ that make element $u'_{qj}$ equal to zero. Using the expression above we can find that if $\phi = \phi_k = \pi + \varphi_{pj} - \varphi_{qj}$ and $\theta = \theta_k = arctan\left(-\frac{r_{qj}}{r_{pj}}\right)$, the element $u'_{qj}$ is equal to zero.

**Table S1.** Cartesian coordinates of the molecular systems use in this study.

| Molecule | Coordinates |
|---|---|
| H$_4$ square | H  0.000000000  0.000000000  0.000000000 |
|  | H  0.000000000  0.000000000  1.100000000 |
|  | H  0.000000000  1.100000000  0.000000000 |
|  | H  0.000000000  1.100000000  1.100000000 |
| C$_2$H$_6$ (dR = uniform grid of 23 points over the interval [–0.2, 2.0]) | C  0.000000000  0.000000000 -0.764994000 |
|  | H  1.019690000  0.000000000 -1.164250000 |
|  | H -0.509845000 -0.883078000 -1.164250000 |
|  | H -0.509845000  0.883078000 -1.164250000 |
|  | C  0.000000000  0.000000000  0.764994000 + {dR} |
|  | H  0.509845000  0.883078000  1.164250000 + {dR} |
|  | H -1.019690000  0.000000000  1.164250000 + {dR} |
|  | H  0.509845000 -0.883078000  1.164250000 + {dR} |

| | |
|---|---|
| C$_2$H$_4$ (dR = uniform grid of 27 points over the interval [–0.2, 1.0]) | C -0.156895360 0.000000000 -0.705755575 |
| | H -0.156895360 0.923341000 -1.278705575 |
| | H -0.156895360 -0.923341000 -1.278705575 |
| | C -0.156895360 0.000000000 0.624402425+{dR} |
| | H -0.156895360 0.923341000 1.197352425+{dR} |
| | H -0.156895360 -0.923341000 1.197352425+{dR} |
| C$_6$H$_6$ | C -1.396328944 0.000000000 0.000000000 |
| | C -0.698164472 -1.209256337 0.000000000 |
| | C -0.698164472 1.209256337 0.000000000 |
| | H -1.241338149 -2.150060743 0.000000000 |
| | H -1.241338149 2.150060743 0.000000000 |
| | C 0.698164472 -1.209256337 0.000000000 |
| | C 0.698164472 1.209256337 0.000000000 |
| | H 1.241338149 -2.150060743 0.000000000 |
| | H 1.241338149 2.150060743 0.000000000 |
| | C 1.396328944 0.000000000 0.000000000 |
| | H 2.482676297 0.000000000 0.000000000 |
| | H -2.482676297 0.000000000 0.000000000 |



\end{document}